\documentclass[reprint, twocolumn, amsmath, amssymb, superscriptaddress]{revtex4-2}

\usepackage{graphicx}
\usepackage{dcolumn}
\usepackage{amsmath}

\usepackage[usenames,dvipsnames]{xcolor} 
\usepackage{hhline}
\usepackage{array}
\newcolumntype{P}[1]{>{\centering\arraybackslash}p{#1}}
\usepackage{textcomp}
\usepackage{comment}
\usepackage[normalem]{ulem}

\begin{document}

\title{Pushing single atoms near an optical cavity}

\author{Dowon Lee}
\affiliation{Department of Electrical Engineering, Pohang University of Science and Technology (POSTECH), 37673 Pohang, Korea}
\author{Taegyu Ha}
\affiliation{Department of Electrical Engineering, Pohang University of Science and Technology (POSTECH), 37673 Pohang, Korea}
\author{Donggeon Kim}
\affiliation{Department of Electrical Engineering, Pohang University of Science and Technology (POSTECH), 37673 Pohang, Korea}
\author{Keumhyun Kim}
\affiliation{Department of Electrical Engineering, Pohang University of Science and Technology (POSTECH), 37673 Pohang, Korea}
\author{Kyungwon An}
\affiliation{Department of Physics and Astronomy \& Institute of Applied Physics, Seoul National University, 08826 Seoul, Korea}
\author{Moonjoo Lee}
\email{moonjoo.lee@postech.ac.kr}
\affiliation{Department of Electrical Engineering, Pohang University of Science and Technology (POSTECH), 37673 Pohang, Korea}

\date{\today}

\begin{abstract}
Optical scattering force is used to reduce the loading time of single atoms to a cavity mode. 
Releasing a cold atomic ensemble above the resonator, we apply a push beam along the direction of gravity, offering fast atomic transport with narrow velocity distribution.
We also observe in real time that, when the push beam is illuminated against gravity, single atoms slow down and even turn around in the mode, through the cavity-transmission measurement.
Our method can be employed to make atom-cavity experiments more efficient. 
\end{abstract}

\maketitle

\section{Introduction}

Coherent atom-photon interaction is a cornerstone in quantum optics and quantum information science~\cite{Haroche06}.
When atoms are coupled to a high-$Q$ resonator, the atoms can interact with the cavity field at single- or few-photon levels.
This enabled important observations like quantized Rabi oscillation~\cite{Brune96}, nonclassical cavity-field states~\cite{Rempe91, Deleglise08}, and superradiance and superabsorption~\cite{Baumann10, Kim2018, Yang2021}.
Coherent interaction is also essential for exploiting the atom-cavity setting for quantum networks.
For instance, the demonstration of single-photon generation~\cite{Hijlkema07, Kang2011}, quantum memory~\cite{Specht11}, and gate operation~\cite{Hacker2016} lay the foundation for constructing a cavity-based quantum network node~\cite{Reiserer2015, Daiss2021, Krutyanskiy2023}.
Recently, the cavity quantum electrodynamics (QED) research becomes more diverse: The coherent interaction in conjunction with dissipation made it possible to explore the non-Hermitian physics~\cite{Choi2010, Kim2023a}, and a combination with the tweezer-array technique enabled the superresolution imaging~\cite{Deist2022} and collective atom-photon interaction~\cite{Liu2023}.

All cavity QED experiments start with loading atoms to a cavity mode. 
For atomic ensembles~\cite{Black03, Bohnet12, Periwal2021} or trapped ions~\cite{Herskind09, Stute12, Begley16, Lee2019, Schupp2021} coupled to centimeter-scale cavities, the atoms were trapped at the cavity mode directly.
However, in many experiments for single or few neutral atoms with small mode volumes, specific techniques have been required for delivering atoms from a source position to the resonator.
Atomic beams were often used to strongly couple atoms to the cavity mode~\cite{Thompson92, Lee2014, Kim2018, Yang2021, Kim2022}.
In order to achieve longer atom-cavity interaction times, laser-cooled atoms were released from a magneto-optical trap (MOT)~\cite{Mabuchi96b, Zhang2011} or Bose-Einstein condensate~\cite{Oettl2005} so that the atoms fall through the cavity by gravity. 
The atoms were also loaded to the cavity from an atomic fountain underneath~\cite{Muenstermann99a, NisbetJones2013}, transported via a moving standing-wave field~\cite{Khudaverdyan2008}, and guided by a magnetic trap~\cite{Gehr2010} or an optical dipole force~\cite{Nussmann05a, Leonard2014, Yang2019}.
While all such approaches have worked, it would always be necessary to load atoms in faster and more stable manners: This not only reduces the time duration of an experimental sequence, but also enables more efficient operation of the system.

Here, we facilitate the loading of single atoms to an optical resonator. 
After a MOT is released above the cavity, we illuminate a push beam to the falling atoms along the direction of gravitational force.  
This optical pushing force decreases the loading time as well as reduces the spread of the atomic velocity distribution. 
Moreover, when the push beam is applied along the opposite direction of gravity, the real-time cavity transmission shows that the single atoms are decelerated, turn around, and leave the cavity mode eventually. 
Our experimental data are explained by solving the It$\hat{\textrm{o}}$ stochastic differential equations, showing atomic trajectories dominated by gravity, the push beam, and diffusion by the cavity fields.

\section{Results}

\subsection{Experimental setup}

\begin{figure} [!t] 
	\includegraphics[width=3.3in]{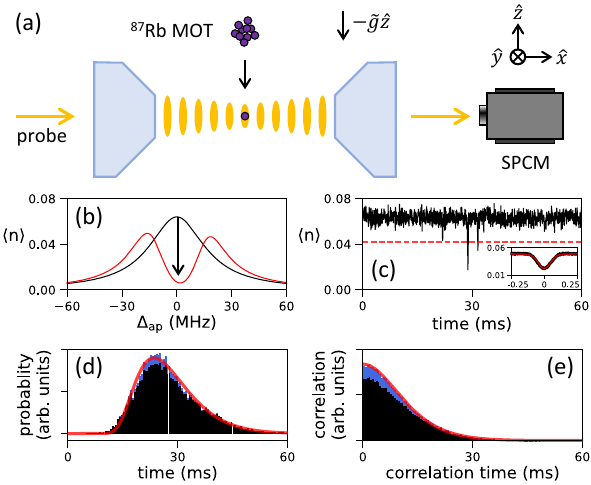} 
	\caption{
		(a) Scheme of the experiment. 
		Laser-cooled atoms fall through the cavity by gravity.
		Cavity transmission of the probe field is measured with a single photon counting module (SPCM).
		Magneto-optical trap (MOT) at the origin.
		(b) Cavity mean photon number $\langle n \rangle$ without an atom (black) and with a maximally coupled single atom (red).
		Cavity transmittance decreases upon the atomic transit (black arrow).
		Atom-probe detuning is $\Delta_\textrm{ap}$.
		(c) Experimentally measured cavity transmission of atomic transits.
		Horizontal red dashed line indicates a threshold for atomic events.
		Bin time is $50$~$\mu$s.
		(inset) Zoomed-in view of the atomic transit event. 
		Average of 1,000 cavity transmissions (black) and theoretical result (red).
		Bin time is $1$~$\mu$s.
		(d) Measured ``on" atomic arrival time distribution (black), multi-atom reconstructed distribution (blue), and fitting with Eq.~\eqref{eqref:histogram} (red).
		Bin time is $500$~$\mu$s.
		(e) Second-order correlation function of ``on" atomic events (black) and with reconstructed multi-atom events (blue). 
		Comparison with $g^{(2)}(\tau)$ (red).
		We use $\sim$~20,000 atomic events for the black bars of (d) and (e).
		}
	\label{fig:setup}
\end{figure}

The experimental setup is presented in Fig.~\ref{fig:setup}(a)~\cite{Kim2021, Lee2022}. 
Our $^{87}$Rb MOT is of $\sim10^{4}$ atoms above the cavity mode. 
The resonator consists of mirrors with a radius of curvature of 10~cm with different transmission losses of 2~and 200~ppm, respectively.
The cavity parameters are $(g_{0}, \kappa, \gamma) = 2 \pi \times (16.02(1), 18.6(5), 3.033(5))$~MHz, where $g_{0}$ is the maximum atom-cavity coupling constant, $\kappa$ is the cavity decay rate, and $\gamma$ is the atomic decay rate.
The cavity mode waist $w_{0}$ is $26.198(8)$~$\mu$m with a length of 151.686(2)~$\mu$m. 
Our atom-cavity system operates in the intermediate coupling regime with a critical photon number of $\gamma^{2}/(2g_{0}^{2})=0.01793(6)$ and cooperativity parameter of $g_0^2/(2\kappa\gamma)=2.28(6)$.
The quantization axis is defined by a magnetic field of $0.4$~G along the cavity axis.
For all experiments in this paper, the cavity resonance frequency $\omega_{\textrm{c}}$ is identical with the atomic transition frequency $\omega_{\textrm{a}}$ (from $5^{2}$S$_{1/2}|F=2, m_{F}=+2\rangle$ to $5^{2}$P$_{3/2}|F'=3, m_{F}=+3\rangle$) at 780~nm. 
The probe laser has the frequency $\omega_{\textrm{p}}$ same as $\omega_{\textrm{a}}$ and $\omega_{\textrm{c}}$, with the $\sigma^{+}$ polarization.
The cavity frequency is stabilized to a weak $\sigma^{-}$-polarized 788-nm laser, yielding a maximum ac Stark shift of $\Delta_{\textrm{st}}=-1.0(2)$~MHz ($47(9)$~$\mu$K)~\cite{Cho2023, SI-OSA}.
This energy shift gives rise to a conservative force whose scale is much smaller than a kinetic energy of free-falling atom of $\sim 470$~$\mu$K: We take $\Delta_{\textrm{st}}$ into account in the numerical simulations of Fig.~\ref{fig:setup}(b), Fig.~\ref{fig:setup}(c) inset, Fig.~\ref{fig:push_above}(c) inset, and Fig.~\ref{fig:push_below}.

\subsection{Reference experiment}

\subsubsection{Atom detection}

A reference experiment is described. 
We begin with loading atoms to the MOT for $70$~ms. 
Switching the quadrupole magnetic field off, sub-Doppler cooling is done for $1$~ms with the cooling laser whose frequency is red-detuned by $59$~MHz from $\omega_{\textrm{a}}$. 
The cooling laser is then turned off at $t_{0}=0$~ms; afterwards the atomic motion is governed by gravity. 
The cavity is driven with a probe laser such that the cavity mean photon number $\langle n \rangle=0.06(1)$ on resonance at the bare cavity condition. 
As shown in Fig.~\ref{fig:setup}(b), strong atom-cavity interaction makes the cavity transmission decrease significantly when a single atom is coupled to the cavity, allowing us to identify the atomic transit. 
The red theoretical line in Fig.~\ref{fig:setup}(b) is obtained via numerically solving the master equation with our experimental parameters for a single atom maximally coupled to the cavity~\cite{SI-OSA}.
The slight asymmetry in the vacuum Rabi spectrum is due to $\Delta_{\textrm{st}}$.

Given cavity transmission data in Fig.~\ref{fig:setup}(c), we judge an atomic arrival only when the minimum value of a dip is measured below a threshold value $\langle n_{\textrm{th}} \rangle =0.045$.
The choice of $\langle n_{\textrm{th}} \rangle$ is done through $\langle n_{\textrm{th}} \rangle = \langle n \rangle -4\sigma_{ \textrm{th} }$, where $\sigma_{\textrm{th}}$ is the standard deviation of measured photon numbers. 
This reduces the probability that we count a shot noise of $\langle n \rangle$ as a false atomic arrival below $10^{-2}$~\cite{SI-OSA}; if we observe a dip with a minimum value smaller than $\langle n_{\textrm{th}} \rangle$, we define the atomic arrival time as the time when the minimum value is detected.

In Fig.~\ref{fig:setup}(c), we find three dips with different minimum values.
We attribute this difference to the position-dependent coupling along the transverse direction. 
The decrease is more pronounced when an atom falls near the mode center $(y\approx0)$ than around the shoulder or tail of the mode $(|y| \gtrsim w_{0})$. 
Note that, the initial position difference along the $x$ axis, between nodes and antinodes, does not affect the minimum value of the dip.
It is because the atom-cavity coupling constant is averaged by the diffusive atomic motion along the cavity axis~\cite{Doherty00b}.

In our atom counting, only ``on" or ``off" of the atomic events are detected, because it is not possible to tell multi-atom events, in which more than two atoms traverse the cavity in $T_{\textrm{int}}$, from individual dips in Fig.~\ref{fig:setup}(c).
We define $T_{\textrm{int}}$ as an approximate atom-cavity interaction time of $2w_{0}/v_{z} \simeq 175$~$\mu$s, with an atomic velocity at the cavity mode of $v_{z}\simeq 0.3$~m/s. 
Therefore, when a minimum value of a dip, smaller than $\langle n_{\textrm{th}} \rangle$, is detected at $t$, we count that ``one" atom traverses the cavity between $(t-T_{\textrm{int}}/2)$ and $(t+T_{\textrm{int}}/2)$~ --- This event is defined as ``on'' in our scheme, and for other times we assign ``off'': The effective deadtime of our atom counter corresponds to $T_{\textrm{int}}$.
In this way we plot a histogram of atomic arrivals, shown as black bars in Fig.~\ref{fig:setup}(d).

\begin{figure*} [!t]
	\includegraphics[width=6.0in]{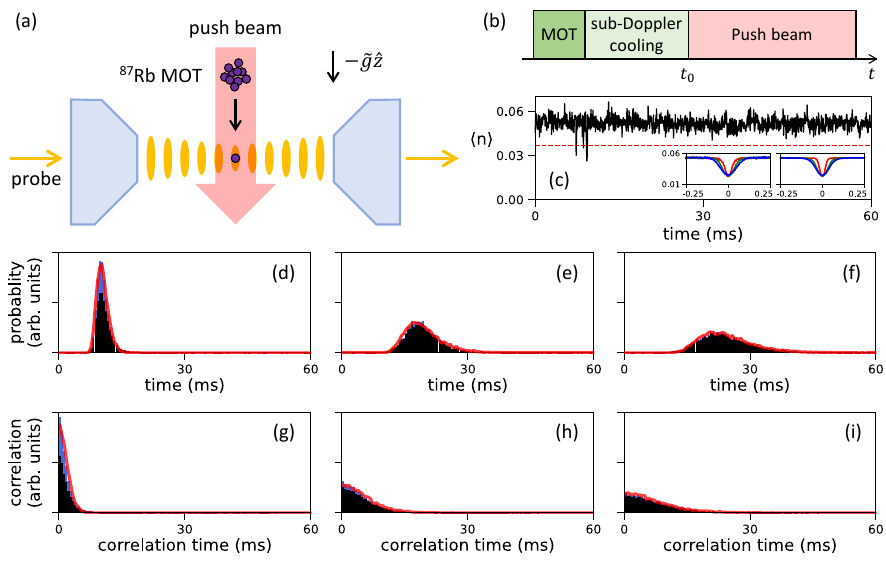} 
	\caption{
	(a) Push beam from above.
	(b) Experimental sequence. 
	Push beam is switched on at $t_0$. 
	(c) Experimentally measured cavity transmission when single atoms traverse the cavity with the push beam at $\Delta_{\textrm{aps}} / (2\pi)=0$~MHz.
	Horizontal red dashed line denotes a threshold for counting the atomic transits.
	(inset, left) Zoomed-in view of the atomic transit events.
	Average of 1,000 cavity transmissions for $\Delta_\textrm{aps} / (2\pi) = 0$ (red), $-10$ (green), and $-20$ (blue) MHz, respectively.
	(inset, right) Theoretical results in given experimental parameters.
	(d)--(f) Distribution of atomic arrival times for $\Delta_\textrm{aps} / (2\pi) = 0, -10,$ and $-20$~MHz, respectively.
	$\sim$~9,000 atomic transits are used to construct each black histogram.
	Histogram with reconstructed multi-atom events is shown in blue. 
	Red line is from Monte Carlo simulation. 
	(g)--(i) Second-order correlation function of the arrival times for $\Delta_\textrm{aps}/(2\pi) = 0, -10,$ and $-20$~MHz, respectively.
	Color code is same as (d)--(f).
	}
	\label{fig:push_above}
\end{figure*}

Furthermore, we reconstruct the multi-atom events that we cannot measure directly from the atomic transit data.
The results are presented with blue bars of Fig.~\ref{fig:setup}(d), $P_\textrm{arr}(t_{i})$, giving full statistics of atomic arrivals.
The reconstruction method is explained in Sec.~\ref{sec:discussion} with a complete description in Ref.~\cite{SI-OSA}.

\begin{figure*} [t]
	\includegraphics[width=6.0in]{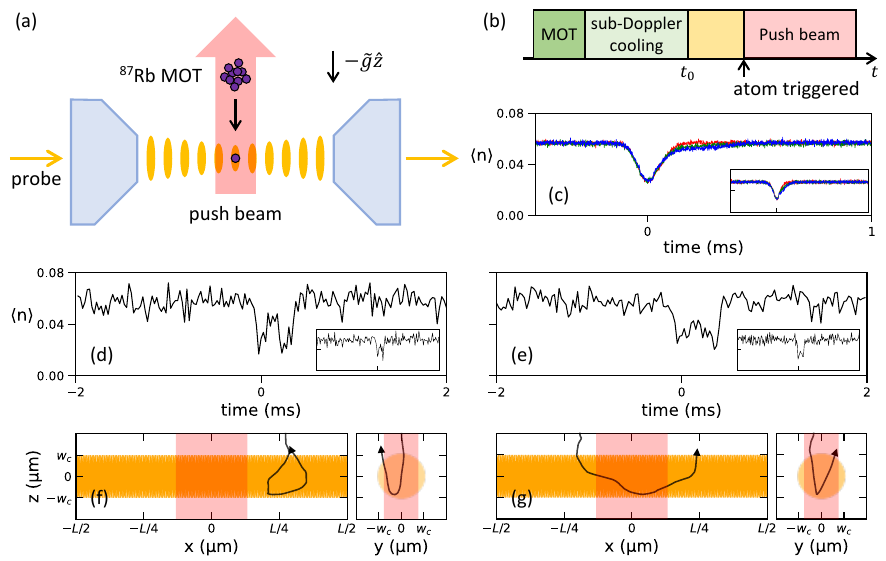} 
	\caption{
		(a) Push beam from below.
		(b) Experimental sequence.
		Push beam is switched on once an atomic transit is triggered. 
		(c) Average of $500$ cavity transmissions for $s = 1.0 \cdot 10^{-1}$ (red), $1.0$ (green), and $1.9$ (blue), respectively.
		(inset) Average of calculation results in given experimental parameters.
		Single-shot measurements of cavity transmission with the push beam below at (d) $s=1.5$ and (e) $s=0.8$. 
		(inset) Calculation results showing similar transmission behaviors. 
		(f), (g) Calculated atomic trajectories for explaining ``double-dip" features of (d) and (e). 
		Vertical red zone indicates the push beam and horizontal orange one denotes the cavity mode. 
	}
	\label{fig:push_below}
\end{figure*}

\subsubsection{Analysis of atomic distribution}

From $P_\textrm{arr}(t_{i})$, we obtain a mean atomic arrival time of $\bar{t}_{\textrm{arr}} = \sum_{i} t_{i} P_\textrm{arr}(t_{i}) = 27$~ms with a standard deviation of $\Delta t_{\textrm{arr}} = \sqrt{ \sum_i t_{i}^{2} P_\textrm{arr}(t_{i}) - \left( \sum_{i} t_{i} P_\textrm{arr}(t_{i}) \right)^2 } = 8$~ms.
The histogram is fitted with

\begin{equation}
	P(t) = c \cdot \sqrt{ \frac{m}{2\pi k_{\textrm{B}}T} }
	\left( \frac{\frac{\tilde{g}t^{2}}{2} + d }{t^{2}} \right)
	e^{-\frac{m}{2k_{\textrm{B}}Tt^2} \left( \frac{\tilde{g}t^2}{2} - d \right)^2  } 
	\label{eqref:histogram}
\end{equation}

\noindent
where $c$ is a coefficient, $m$ is the atomic mass of $^{87}$Rb, $k_{\textrm{B}}$ is the Boltzmann constant, $T$ is the atomic temperature at $t_{0}$, $d$ is the distance between the MOT and cavity center, and $\tilde{g}$ is the gravitational acceleration~\cite{Yavin2002, SI-OSA}.
Fitting the histogram in Fig.~\ref{fig:setup}(d) gives $T=83(3)$~$\mu$K and $d=4.80(7)$~mm.

\subsubsection{Correlation function}

The second-order correlation function, $g^{(2)}(\tau)$, of atomic arrival events is presented in Fig.~\ref{fig:setup}(e).
We show not only the histogram of the ``on" events, but also that with reconstructed multi-atom occasions. 
The process of the latter is given as follows. 
Through the reconstruction process, we obtain the total number of reconstructed atomic events for every $i^{\textrm{th}}$ bin. 
We randomly distribute these events to each bin of all sequences.
The correlations are obtained from the measured ``on" atomic events with these reconstructed events. 
More details are provided in Ref.~\cite{SI-OSA}.

In Fig.~\ref{fig:setup}(e), the measured correlation is compared with $g^{(2)}(\tau) = \langle P(t)P(t+\tau) \rangle / \langle P(t) \rangle^{2}$, showing an agreement between  $g^{(2)}(\tau)$ and the correlation with the reconstructed distribution. 
We also note that the atom statistics show clear bunching behavior of our ``pulsed" atomic injection; the period of the atomic pulse is the duration of one experimental sequence.

\subsection{Push beam from above}

We discuss next experiment in which the push beam is applied from above, i.e., along the direction of gravity (Fig.~\ref{fig:push_above}(a)).
The experimental sequence is given in Fig.~\ref{fig:push_above}(b). 
After the sub-Doppler cooling is finished at $t_{0}=0$~ms, the push beam is switched on, transporting atoms from the atomic source position to the cavity mode. 
The atom-push beam detuning is $\Delta_\textrm{aps}/(2\pi) = \omega_\textrm{a} - \omega_\textrm{ps}$, changing from $0$ to $-20$~MHz with the frequency of the push beam $\omega_\textrm{ps}$.
The waist of the push beam is $\sim20$~$\mu$m with a saturation parameter of $s=I/I_\textrm{s} \simeq 1.0 \cdot 10^{-3}$, where $I$ is the intensity of the push laser and $I_\textrm{s}=1.67$~mW/cm$^{2}$ is the saturation intensity of the transition from $5^{2}$S$_{1/2}|F=2\rangle$ to $5^{2}$P$_{3/2}|F'=3 \rangle$.
The repump laser field (resonant on the transition from $5^{2}$S$_{1/2}|F=1\rangle$ to $5^{2}$P$_{3/2}|F'=2 \rangle$, same as the field used in MOT) overlaps and co-propagates with the push beam.

The cavity transmission in Fig.~\ref{fig:push_above}(c) presents how the atomic transit is affected in this configuration. 
It is notable that the atoms arrive at the cavity mode much faster than the cases of Fig.~\ref{fig:setup}(c) --- This is the major impact of the push beam.
In Figs.~\ref{fig:push_above}(d)--(f), we show the histograms of arrival times for $\Delta_{\textrm{aps}}/(2\pi)=0, -10,$ and $-20$~MHz, in the order of descending optical scattering force. 
We also reconstruct multi-atom events following the same procedure used in Fig.~\ref{fig:setup}(d).
Three features are pointed out here. 
First, as the optical force increases, the net speed of the atom enhances, and thus the arrival time decreases for stronger optical force.
We obtain $\bar{t}_{\textrm{arr}}(\Delta t_{\textrm{arr}})= 11(2), 21(5),$ and $25(7)$~ms, respectively for the reconstructed data of Figs.~\ref{fig:push_above}(d)--(f).
Second, when more optical force is applied, the width of the velocity distribution decreases. 
This is because faster (slower) atoms are exposed to the optical force for shorter (longer) interaction times so that the overall velocities converge to a central value.
Finally, as shown in insets of Fig.~\ref{fig:push_above}(c), we observe that the transmission dip becomes narrow as the atomic velocity increases, agreeing with the simulation results.

We quantitatively understand this observation using

\begin{equation}
	\label{eq:force}
	\frac{dv_{z}}{dt} = -\tilde{g} - \frac{ \hbar k }{m}  \frac{\gamma \cdot s}{ 1 + s + \left( (\Delta_{\textrm{aps}} -kv_{z}) / \gamma \right)^{2} },
\end{equation}

\noindent
with the atomic velocity along the $z$ direction $v_{z}$, gravitational acceleration constant $\tilde{g}$, Planck constant $\hbar$, and wavevector of the push beam $k$. 
We perform MC simulations that calculate the histogram of atomic arrival times~\cite{SI-OSA}. 
Assuming the atom as a classical particle, we numerically calculate the atomic acceleration and velocity through Eq.~\eqref{eq:force}, and the associated position in a three-dimensional space. 
The results are shown in Figs.~\ref{fig:push_above}(d)--(f), with $\bar{t}_\textrm{arr} (\Delta t_{\textrm{arr}})=10(1), 20(5),$ and $25(7)$, respectively, agreeing with the experimental data. 
Note that we make use of $s=0.66 \cdot 10^{-3}, 1.1 \cdot 10^{-3}$, and $1.0 \cdot 10^{-3}$ for each simulation to find the agreement. 
The difference of $s$ between the experiment and theory would be attributed to slight drift of the MOT position over the whole measurements.

Summarizing the impact of the push beam, we compare the result of Fig.~\ref{fig:push_above}(d) with that of the free-falling case of Fig.~\ref{fig:setup}(d).
The atomic loading time from the MOT to the cavity mode $(\bar{t}_{\textrm{arr}})$ decreases by a factor of $2.5$, with a fourfold reduction of the arrival-time distribution $(\Delta t_{\textrm{arr}})$.

We then obtain $g^{(2)}(\tau)$ of the atomic arrival events under the push beam. 
The results are presented in Figs.~\ref{fig:push_above}(g)--(i).
The atom statistics reveal a bunching behavior; stronger optical force leads to more bunched atomic distribution. 
This is also consistent with the results of MC simulations.

When the push beam is shined to a MOT with similar atom number with that of Fig.~\ref{fig:setup}, we observe that the atomic flux to the cavity mode increases significantly, because the atoms in the MOT would arrive at the mode with a reduced arrival time distribution.
In such configuration it is frequently observed that the individual dips overlap, making the data analysis complicated. 
Therefore, for the experiments with the push beam, the measurements are conducted with smaller atom number in the MOT so that in most of the cases we observe separate individual dips, like Fig.~\ref{fig:push_above}(c).

\subsection{Push beam from below}

We continue to the experiments where the push beam is illuminated from below (Fig.~\ref{fig:push_below}(a)).  
Fig.~\ref{fig:push_below}(b) sketches the experimental sequence.
Like aforementioned experiments, we monitor atomic arrivals via the real-time cavity-transmission measurement. 
When a transmission signal decreases below $\langle n_{\textrm{th}} \rangle$, we turn the push beam on such that the optical scattering force is applied against the direction of gravity. 
The frequency of the push beam is $\Delta_{\textrm{aps}}/(2\pi)=0$~MHz, the waist is similar with that from above, and the beam is overlapped with a co-propagating repumping laser.

One experimental result is shown in Fig.~\ref{fig:push_below}(c). 
Each colored data corresponds to an averaged cavity transmission of atomic transits for different $s$. 
It is notable that, as the optical force increases, the dip becomes more asymmetric: The second half of the dip exhibits `tail-like' behavior as contrary to the first half.
This feature reflects that the atomic motion is under the impact of the push beam.
The atoms are decelerated by the scattering force, and stay in the cavity for longer interaction times. 
Eventually the continuous optical force changes atomic propagation to the same direction of the force, leading the atom to leave the cavity mode.

Such atomic motion is more clearly manifested in Figs.~\ref{fig:push_below}(d)-(g).
Interestingly, while monitoring the atomic transits, we often observe that the dips appear twice with the push beam below, like Figs.~\ref{fig:push_below}(d) and (e). 
It is because the atom passes through the cavity center (along the $z$ direction) twice: Falling through the cavity, the atom crosses the center, and turns its propagation direction below $z < -d-w_{0}$. 
The atom passes through the cavity center again when going up against gravity.

In order to ``visualize'' this scenario, we solve the It$\hat{\textrm{o}}$ stochastic differential equations to simulate the atomic trajectories~\cite{SI-OSA}. 
We consider not only the scattering force of the push beam, but also the dipole force of the cavity-frequency stabilization laser, the cavity-field induced friction force, fluctuations of the dipole force, and the effect of recoil kick caused by the spontaneous emission~\cite{Doherty00b, SI-OSA}.
We find similar features with the experimental data through the feasible cavity transmissions in the insets of Figs.~\ref{fig:push_below}(d) and (e). 
Also, the corresponding atomic trajectories are depicted in Figs.~\ref{fig:push_below}(f) and (g), explaining the ``double-dip'' feature of the measurement data.

\section{Discussion}
\label{sec:discussion}
First, we describe the method for reconstructing multi-atom events. 
In Fig.~\ref{fig:setup}(d), each black bar gives the information about the mean number of ``on'' events (probability times number of sequences) at the $i^{\textrm{th}}$ bin.
We then assume that the actual atom number distribution at the $i^\textrm{th}$ bin, over all sequences, is Poissonian. 
Next, it is pointed out that the obtained ``on'' events include all probabilities except $p_{i}(0)$, accordingly quantifying $p_{i}(0)$ of the Poissonian distribution.
This determines the actual mean atom number of the $i^{\textrm{th}}$ bin, $\langle N_{i}^{\textrm{a}} \rangle$, through the relation $p_{i}(0)= e^{-\langle N_{i}^{\textrm{a}} \rangle}$~\cite{SI-OSA}.
In such manner we obtain $\langle N_{i}^{\textrm{a}} \rangle$, which identifies the probabilities of all multi-atom events in the $i^\textrm{th}$ bin, allowing us to reconstruct the full atomic arrival time distribution $P_{\textrm{arr}}(t_{i})$ (blue bars in Fig.~\ref{fig:setup}(d)).
Complete description is provided in Ref.~\cite{SI-OSA}; we stress that, differently from similar atom-cavity experiments~\cite{Hood98, Zhang2011}, we reconstruct all the multi-atom events that offer full statistics of the atomic transits.

Second, we remark the total number of atoms crossing the cavity mode per release $\langle N_{\textrm{cav}} \rangle$ and the effect of a red-detuned laser, given the push beam from above.
Our MC simulation shows that, when the atoms of the same number are in a MOT, $\langle N_{\textrm{cav}}\rangle$ at $\Delta_{\textrm{aps}}/(2\pi)=0$~MHz increases by a factor of $\sim 5$ than that of the free-fall case.
In the experiment it is not possible to measure this enhancement because we could not accurately count $\langle N_{\textrm{cav}} \rangle$ due to the overlap of the dips in $T_{\textrm{int}}$; if we would perform the experiment with a MOT with much less atoms, we would examine it quantitatively.
Regarding the red-detuned push beam, as estimated from Eq.~\eqref{eq:force}, the push-beam force as a function of the laser frequency is symmetric with respect to $kv_{z} / (2\pi) \simeq 0.4$~MHz for $v_{z} \simeq 0.3$~m/s of a free-falling atom. 
Therefore, we expect that the atomic histogram would be very similar for the two push beams of the same $|\Delta_{\textrm{aps}}|$ (particularly when $|\Delta_{\textrm{aps}}| \gg kv_{z}$), also observed in our measurements.

Finally, we envisage an atomic loading to the cavity in which the two push beams, from above and below, are utilized sequentially. 
One can decrease the transport time from the MOT position to the cavity mode through the push beam above. 
As the push beam above is switched off, the push beam below is applied until the atom ``stops" or becomes sufficiently slow in the cavity mode, by dissipating the atomic kinetic energy gained through gravity and the first push beam.
The atom can then be trapped in the cavity mode using a far-off resonant laser. 
This technique would bring about fast atomic loading, with highly probable capture of single atoms into an intracavity dipole trap.

\section{Conclusion}

In conclusion, we have observed the impact of the push beam to single slow atoms near an optical cavity. 
The atomic dynamics are controlled in two ways.
First, via applying an optical scattering force along the direction of gravity, the atomic center velocity increases with reduced spread of the velocity distribution: This allows us to achieve the fast transport of single atoms from the source position to the cavity mode.
Second, when the push beam is illuminated from below, we observe that the atoms slow down and the optical force even turns the atomic propagation direction around. 
Our work represents not only \textit{in-situ} monitoring of the effect of the push beam onto a single atom, but also a useful technique for optical cavity-QED experiments.

\begin{acknowledgements}
We thank D.~Cho for providing us with a numerical tool that calculates the ac Stark shift. 
This work has been supported by  BK21 FOUR program and Educational Institute for Intelligent Information Integration, National Research Foundation (Grant No.~2019R1A5A1027055 and Grant NO.~2020R1I1A2066622), Institute for Information \& communication Technology Planning \& evaluation (IITP, Grant No.~2022-0-01040), Samsung Science and Technology Foundation (SRFC-TC2103-01) and Samsung Electronics Co., Ltd. (IO201211-08121-01).
K.~An was supported by the Korea Research Foundation (Grant No.~2020R1A2C3009299).
\end{acknowledgements}

Our data are available at https://doi.org/10.5281/zenodo.10829540.

\bibliographystyle{apsrev4-2}
\bibliography{bibliography}

\clearpage
\onecolumngrid
\appendix

\section{Hamiltonian and master equation}
\label{sec:Hamiltonian}

We describe the theoretical calculation for Fig.~1(b), Fig.~1(c) inset, Fig.~2(c) inset, and Fig.~3.
The Hamiltonian describing the atomic internal state, cavity field, and probe laser is

\begin{equation}
	H_{\textrm{tot}} = \hbar \left( \omega_{\textrm{a}} + \Delta_{\textrm{st}}(\vec{r}) \right) \sigma_{\textrm{e}}  +  \hbar\omega_{\textrm{c}} a^{\dagger}a
	+ \hbar g(\vec{r}) ( a^{\dagger}\sigma_{\textrm{ge}} + \textrm{h.~c.} )
	+  \hbar \eta ( a^{\dagger} e^{-i\omega_{\textrm{p}}t} + \textrm{h.~c.} ),
	\label{eq:Htot}
\end{equation}

\noindent
where $\hbar$ is the Planck constant, $\omega_{\textrm{a}}$ is the transition frequency between $|$g$\rangle (\equiv 5^{2}$S$_{1/2}|F=2, m_{F}=+2\rangle$) and $|$e$\rangle (\equiv 5^{2}$P$_{3/2}|F'=3, m_{F}=+3\rangle )$, $\Delta_{\textrm{st}}(\vec{r})$ denotes the position-dependent ac Stark shift induced by the cavity-frequency stabilization laser, $\sigma_{\textrm{e}}$ is the projection operator onto the state $|\textrm{e}\rangle$, $\omega_{\textrm{c}}$ is the cavity frequency, $a^{\dagger}(a)$ is the cavity photon creation (annihilation) operator, $g(\vec{r})$ is the position-dependent atom-cavity coupling constant, $\sigma_{\textrm{ge}}(=\sigma_{\textrm{eg}}^{\dagger})$ is the transition operator from $|$e$\rangle$~$(|$g$\rangle)$ to $|$g$\rangle$~$(|$e$\rangle )$, $\eta$ is the amplitude of the probe laser, and $\omega_{\rm{p}}$ is the frequency of the probe laser.
The interaction Hamiltonian is obtained by using the relation $H_{\rm{int}}=i \hbar \dot{U}U^{\dagger} + UH_{\rm{tot}}U^{\dagger} $ with the unitary operator $U =\exp \left( i \omega_{\textrm{p}} (a^{\dagger}a + \sigma_{\textrm{e}}) t \right) $.  
After the transformation, we obtain

\begin{equation}
	H_{\textrm{int}} = \hbar \left( \Delta_{\textrm{ap}} + \Delta_{\textrm{st}}(\vec{r}) \right) \sigma_{\textrm{e}}  + \hbar \Delta_{\textrm{cp}} a^{\dagger} a 
	+ \hbar g(\vec{r}) \left( a^{\dagger} \sigma_{\textrm{ge}} + \textrm{h.~c.} \right)
	+ \hbar \eta \left( a^{\dagger} + \textrm {h.~c.} \right),
\end{equation}

\noindent
with $\Delta_{\textrm{ap}} = \omega_{\textrm{a}} - \omega_{\textrm{p}}$ and $\Delta_{\textrm{cp}} = \omega_{\textrm{c}} - \omega_{\textrm{p}}$.
The master equation for the atom-photon density matrix $\rho$ is then expressed as

\begin{align}
	\frac{d\rho}{dt} & = -\frac{i}{\hbar} \left[ H_{\textrm{int}}, \rho \right] + D_{\textrm{at}}[\rho] + D_{\textrm{cav}}[\rho] \nonumber \\ 
	& = -\frac{i}{\hbar} \left[ H_{\textrm{int}}, \rho \right] 
	+  \gamma \left( 2 \sigma_{\textrm{ge}} \rho \sigma_{\textrm{eg}} 
	- \sigma_{\textrm{eg}} \sigma_{\textrm{ge}} \rho - \rho \sigma_{\textrm{eg}} \sigma_{\textrm{ge}}  \right)
	+ \kappa \left( 2a \rho a^{\dagger} - a^{\dagger}a\rho - \rho a^{\dagger}a \right),
	\label{eq:eom}
\end{align}

\noindent
where $D_{\textrm{at(cav)}}[\rho]$ describes the dissipation of the atomic state (cavity field), $\gamma$ is the atomic decay rate from $|\textrm{e}\rangle$ to $|\textrm{g}\rangle$, and $\kappa$ is the cavity decay rate. 
We numerically integrate Eq.~\eqref{eq:eom} in python using QuTip for obtaining the steady state values of the atomic state and cavity mean photon number $\langle n \rangle$.
In Fig.~1(b), the calculation is done as $\omega_{\textrm{p}}$ (accordingly $\Delta_{\textrm{cp}}$) is scanned at a fixed $\Delta_{\textrm{st}}(\vec{r})/(2\pi)= -1$~MHz.
The black line in Fig.~1(b) is obtained with $g(\vec{r})=0$, and red line with $g(\vec{r})=g_{0}$, with the maximum atom-cavity coupling constant $g_{0}$. 
For the simulation results in Fig.~3, we consider the position dependence of $g(\vec{r})$ and $\Delta_{\textrm{st}}(\vec{r})$ with atomic external degrees of freedom (see Sec.~\ref{sec:qmc_sim}).

\section{Determination of threshold photon number}
\label{sec:threshold}

\begin{figure*} [h]
	\includegraphics[width=6.0in]{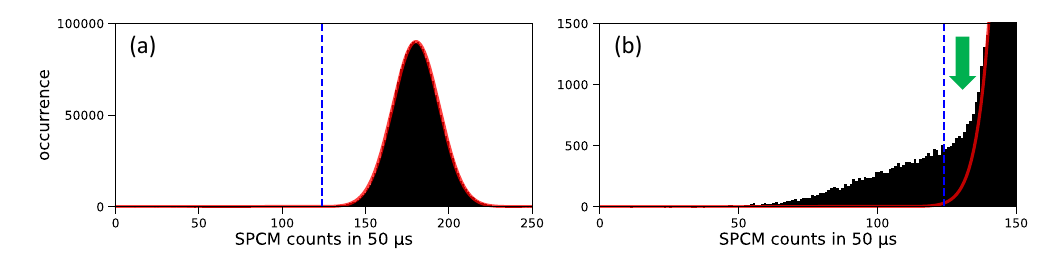} 
	\caption{
		(a) Histogram of measured photon counts.
		Dominant peak is fitted with a Gaussian function (red line). 
		Bin time is $50$~$\mu$s.
		(b) Zoomed-in view of (a).
		Vertical blue dashed line is a threshold of $ \eta \langle n_\textrm{th} \rangle$.
		Green arrow indicates the region of missing atomic signals (see text). 
		Single photon counting module (SPCM).	
	}
	\label{fig:threshold}
\end{figure*}

In Figs.~\ref{fig:threshold}(a) and (b), we plot histograms of single photon counting module (SPCM) counts of $\sim$~2,000 cavity transmission measurements with a bin time of $50$~$\mu$s. 
The mean value of the dominant peak in Fig.~\ref{fig:threshold}(a) originates from the mean photon counts $\bar{C}$ at the bare cavity condition, and the width from the shot noise of the counts. 
This peak is fitted with a Gaussian function $y=c \cdot e^{-(  (x-\bar{C}) / \sigma)^{2}/2  }$ ($c$ is a coefficient) that is of $\bar{C} = 180.29(1)$ and a standard deviation of $\sigma = 14.05(1) \simeq{\sqrt{\bar{C}}}$, reflecting a Poissonian distribution of the measured photon counts. 
In this work, we set a threshold photon number of $\langle n_{\textrm{th}} \rangle = (\bar{C} - 4\sigma) / \eta$ with $\eta = 3.4(7) \cdot 10^{3}$, the conversion factor from the measured photon counts to mean photon number $\langle n \rangle$.
That is, we choose $\langle n_{\textrm{th}} \rangle$ at the level where the measured photon counts are less than $\bar{C}$ by $4\sigma$, indicated by the vertical blue dashed lines in Figs.~\ref{fig:threshold}(a) and (b).

Given $\langle n_{\textrm{th}} \rangle$, we estimate the probability that we would overcount and miss the atomic arrivals. 
The overcounting probability $\eta_{\textrm{over}}$, corresponding to the probability in which we measure a shot noise as an atomic signal, is given by

\begin{equation}
	\eta_{\textrm{over}} = \frac{N_\textrm{shot}}{N_{\textrm{sm}}} = 6.952(9)\cdot 10^{-3},
\end{equation} 

\noindent  
where $N_{\textrm{sm}}$ indicates the number of all counts smaller than $\eta\langle n_{\textrm{th}} \rangle$, and $N_{\textrm{shot}}$ is the number of counts that is below the Gaussian fit among $N_{\textrm{sm}}$.

We also quantify the probability that an atomic signal is lost due to our choice of $\langle n_\textrm{th} \rangle$.
Looking into Fig.~\ref{fig:threshold}(b), we point out that the atomic signals $N_\textrm{la}$ larger than $\eta \langle n_\textrm{th} \rangle$ are lost among the counts above the Gaussian fit (green arrow in Fig.~\ref{fig:threshold}(b)).
The missing probability $\eta_\textrm{miss}$ is thus obtained

\begin{equation}
	\eta_{\textrm{miss}} = \frac{N_{\textrm{la}}}{N_{\textrm{atom}}} = 3.033(4) \cdot 10^{-3},
\end{equation}

\noindent
with the number of all counts above the Gaussian fit $N_\textrm{atom}$, for which we should have measured as ``true" atomic transits.
These lost counts are mostly the cases when the atoms are weakly coupled to the cavity, falling through the tail of the cavity mode $(|y|\gg w_{0})$.

\section{Reconstruction of multi-atom events}
\label{sec:rec_multi_atom_events}

In the cavity transmission data, once a measured photon number at $t$ is smaller than $\langle n_\textrm{th} \rangle$, we count that an atom traverses the cavity between $(t-T_\textrm{int}/2)$ and $(t+T_\textrm{int}/2)$, with an approximate atom-cavity interaction time of $T_\textrm{int} \simeq 175$~$\mu$s, and neglect multi-atom events over this time duration. 
In the following, we describe the estimation of complete zero-, one-, and multi-atom probabilities from the measurement data.

We consider $N_{\textrm{bin}} = T_{\textrm{tot}} / T_{\textrm{bin}}$ time bins with a total measurement time of $T_{\textrm{tot}} = 60$~ms and a bin time of $T_{\textrm{bin}} = 500$~$\mu$s. 
It is assumed that the actual atom number distribution in the $i^{\textrm{th}}$ bin is Poissonian for the $M$ experimental sequences. 
That is, in one sequence, the probability that $k$ atoms arrive in the $i^{\textrm{th}}$ bin is

\begin{equation}
	P_{i}(k) = \frac{ \langle N_{i}^{\textrm{a}} \rangle^{k} e^{-\langle N_{i}^{\textrm{a}}\rangle} }{k!},
	\label{eq:Poisson}
\end{equation}

\noindent
where $\langle N_{i}^{\textrm{a}} \rangle$ is the actual mean atom number in the $i^{\textrm{th}}$ bin ($\langle N_{i}^{\textrm{a}} \rangle < 1$ in our work).
The black bars in Fig.~1(d) correspond to $\langle N_{i}^{\textrm{m}} \rangle / M$, associated with the measured atomic events $\langle N_{i}^{\textrm{m}} \rangle$ in the $i^{\textrm{th}}$ bin over the $M$ measurements $(0 \le \langle N_{i}^{\textrm{m}} \rangle < M)$.
Considering that all multi-atom events are included in individual single-atom transits, we obtain

\begin{equation}
	\frac{ \langle N_{i}^{\textrm{m}} \rangle }{M} =  \sum_{k=1}^{\infty} P_{i}(k) = 1 - P_{i}(0),
	\label{eq:N_over_M}
\end{equation}

\noindent
which immediately yields $P_{i}(0)$.
Since $P_{i}(0)$ is the specific value for a Poissonian distribution of given $\langle N_{i}^{\textrm{a}} \rangle$ (except $\langle N_{i}^{\textrm{a}} \rangle = 1$, and our $\langle N_{i}^{\textrm{a}} \rangle < 1$), we obtain $\langle N_{i}^{\textrm{a}} \rangle$ through numerically solving $P_{i}(0) = e^{-\langle N_{i}^{\textrm{a}} \rangle}$.
This determination of $\langle N_{i}^{\textrm{a}} \rangle$ gives $P_{i}(k)$ for $k\in{[1, \infty]}$ via Eq.~\eqref{eq:Poisson}, allowing us to reconstruct actual one- and multi-atom probabilities.
In other words, while we cannot distinguish the multi-atom events in the atomic transit signals, we indeed measure the probability that \textit{the atom is not detected in the $i^{\textrm{th}}$ bin}: This probability, $P_{i}(0)$, is also that of the Poissonian distribution of actual atomic probabilities, and thus the measurement of $P_{i}(0)$ specifies $\langle N_{i}^{a} \rangle$.

The histogram with reconstructed full atomic events are shown as blue bars in Fig.~1(d): Each blue bar corresponds to $\langle N_{i}^{\textrm{a}} \rangle / M$. 
The reconstructed atom number distribution is fitted well with Eq.~(1), giving the distance between the magneto-optical trap (MOT) and cavity $d$ and the atomic temperature $T$.

\begin{figure*} [h]
	\includegraphics[width=5.0in]{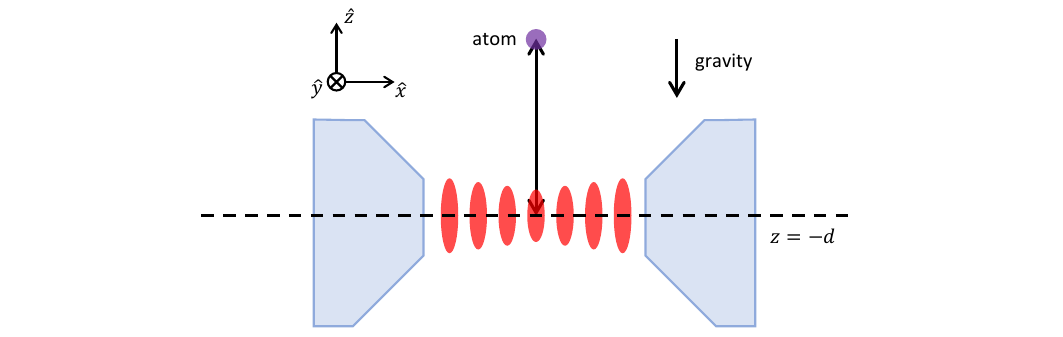} 
	\caption{
		Isotropic atomic cloud at the origin. 
		Cavity center is at $(0, 0, -d)$.
	}
	\label{fig:tof}
\end{figure*}

\section{Histogram of atomic arrival times}
\label{sec:point}

Eq.~(1) is derived based on the approach of Ref.~\cite{Yavin2002}.
We consider that atoms are initially located at the origin (Fig.~\ref{fig:tof}), and neglect the size of the cloud; the atoms start falling at $t=0$.
We aim to calculate the atomic arrival time distribution $P(t)$ at the measurement plane at $z=-d$. 
The Maxwell-Boltzmann isotropic probability distribution for the velocity is given by

\begin{equation}
	N(v) d^{3}v = \left( \frac{m} {2 \pi k_{\textrm{B}} T} \right)^{3/2} \exp \left( - \frac{m ( v_{x}^{2} + v_{y}^2 + v_{z}^{2} ) }{2 k_{\textrm{B}} T }  \right)  d^{3}v,
	\label{eq:MB}
\end{equation}

\noindent
where $m$ is the atomic mass, $k_{\textrm{B}}$ is the Boltzmann constant, and $v_{x,y,z}$ are atomic velocities along the $x, y,$ and $z$ directions, respectively.

We assume a ballistic motion of the atoms under gravity. 
When atoms fall freely, the relations below are obtained

\begin{align}
	x &= v_{x}t \\ 
	y &= v_{y}t \\  
	-d & = v_{z}t - \frac{1}{2} \tilde{g} t^{2},
\end{align}

\noindent
with the gravitational acceleration of $\tilde{g}$.
We perform a transformation of Eq.~\eqref{eq:MB} from the velocity coordinate to the spatial coordinate using the Jacobian determinant

\begin{equation}
	J = \frac{\partial(v_{x}, v_{y}, v_{z})}{\partial(x, y, t)} = 
	\left| \begin{array}{ccc}
		\frac{\partial{v_x}}{\partial{x}} & \frac{\partial{v_x}}{\partial{y}} & \frac{\partial{v_x}}{\partial{t}} \\
		\frac{\partial{v_y}}{\partial{x}} & \frac{\partial{v_y}}{\partial{y}} & \frac{\partial{v_y}}{\partial{t}} \\
		\frac{\partial{v_z}}{\partial{x}} & \frac{\partial{v_z}}{\partial{y}} & \frac{\partial{v_z}}{\partial{t}}
	\end{array} \right|
	=\frac{ \frac{ \tilde{g} t^{2}}{2} + d }{t^{4}},
	\label{eq:Jacobian}
\end{equation}

\noindent
allowing us to obtain $P(t)$ like below: 

\begin{align}
	P(t) & = c \int N(x, y, t) dx dy \\
	& = c \int J \cdot \exp{ \left( -\frac{m}{ 2 k_{\textrm{B} } T t^{2}} \left( x^{2} + y^{2} + \left( \frac{\tilde{g} t^{2} }{2} - d \right)^{2}  \right) \right) } dx dy \\
	& = c \cdot \sqrt{ \frac{m}{2\pi k_{\textrm{B}} T }  }  
	\left( \frac{  \frac{ \tilde{g} t^{2} }{2} + d}{t^{2}} \right)
	\exp \left( -  \frac{m}{2 k_\textrm{B}T t^{2} } \left( \frac{ \tilde{g} t^{2} }{2} - d  \right)^{2}    \right),
	\label{eq:probability}
\end{align}

\noindent 
where $N(x, y, t)$ is the two-dimensional atomic distribution at $t$, and $c$ is a coefficient.

\begin{figure*} [h]
	\includegraphics[width=3.3in]{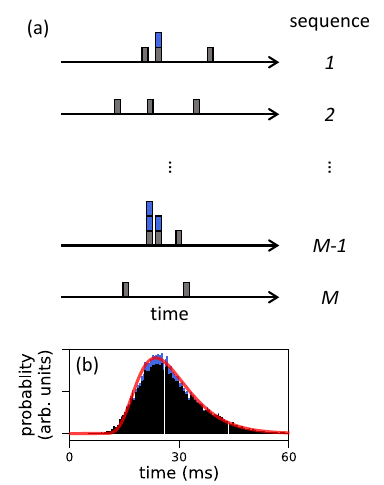} 
	\caption{
		(a) Example time stamps of atomic ``on'' events (dark gray) and reconstructed multi-atom events (blue). 
		(b) Probability distribution of ``on'' events (black) and with multi-atom events (blue). 
		Red line is the fitting with Eq.~\eqref{eq:probability}.
	}
	\label{fig:time_stamp}
\end{figure*}

\section{Correlation function with multi-atom events}

In Fig.~1(e) and Figs.~2(g)-(i), we include multi-atom events in the construction of the correlation function. 
Our strategy for time stamping of these ``reconstructed" events is displayed in Fig.~\ref{fig:time_stamp}(a).
The information we have is the total number of reconstructed events for each $i^\textrm{th}$ bin of the atomic arrival time distribution (Fig.~1(d) and Fig.~\ref{fig:time_stamp}(b)). 
We distribute these events to each bin randomly among the $M$ sequences, like shown as blue bars in Fig.~\ref{fig:time_stamp}(a): These events are used for obtaining the correlation function in Fig.~1(e). 
The identical method is used in Figs.~2(g)-(i).

\section{Monte Carlo simulation with push beam above}
\label{sec:mc_sim}

We describe a Monte Carlo (MC) simulation in which we calculate the histogram of atomic arrival times when the push beam is applied from above (Fig.~2).
In a three-dimensional space, we assume that each atom is initially at the origin with a velocity randomly selected from the MB distribution.
As the atom starts moving, the atomic velocity $v_{z}$ is governed by the optical force from the push beam and the force of gravity, while $v_{x}$ and $v_{y}$ are affected by the recoil kicks from spontaneous emissions. 
Switching the push beam on at $t_{0}=0$, the atomic position $(x_{i}, y_{i}, z_{i})$ is determined as $(x_{i-1} + v_{x, i-1}\Delta t, y_{i-1}+v_{y, i-1}\Delta t, z_{i-1}+v_{z,i-1}\Delta t)$, with a time resolution of $\Delta t = 1~\mu$s.
The impact of the push beam with gravity is expressed as

\begin{align}
	\frac{dv_{z, i}}{dt} & = -\tilde{g} - \frac{\hbar k}{m} \gamma \cdot \sigma_{\textrm{e},i} \nonumber \\
	& =-\tilde{g} - \frac{\hbar k}{m} \frac{\gamma \cdot s}{ 1 + s + \left( (\Delta_{\textrm{aps}} - kv_{z, i})/\gamma \right)^{2} },
\end{align}

\noindent
where $k$ is the wavevector of the cooling laser, $\sigma_{\textrm{e},i}$ is the atomic excitation probability at the step $i$, $s$ is the saturation parameter, and $\Delta_{\textrm{aps}}$ is the atom-push beam detuning. 
We define the saturation parameter $s=I/I_{\textrm{s}}$, where $I_\textrm{s}$ is the saturation intensity of the $5^{2}$S$_{1/2}|F=2\rangle$ to $5^{2}$P$_{3/2}|F'=3 \rangle$ transition; the scattering force of the repumping laser is neglected.
Given $s$ and $\Delta_{\textrm{aps}}$ in Figs.~2(d)--(f) and main text, each histogram is constructed with $\sim10^{4}$ simulated atomic transits.

\section{Calculation of atomic trajectory with push beam below}
\label{sec:qmc_sim}

We carry out a numerical simulation to understand the atomic trajectory when the push beam is applied from below. 
The dynamics is described in the semiclassical regime for atomic external degrees of freedom~\cite{Dalibard1985} (or ``quasiclassical'' regime of Ref.~\cite{Doherty00b}).
In order to calculate the trajectory, we follow the approach of Ref.~\cite{Doherty00b} through solving the It$\hat{\textrm{o}}$ stochastic differential equations

\begin{align}
	d\vec{x} = & \frac{\vec{p}}{m}  dt  \label{eq:Ito_equation_x} \\
	d\vec{p} = & -\hbar g_{0} \langle \Phi \rangle \nabla \psi dt - \frac{\hbar g_{0}^{2}} {m} \chi(\vec{r}) (\vec p \cdot \nabla \psi ) \nabla \psi dt \nonumber \\
	& + \hbar g_{0} \sqrt{2 \xi(\vec{r}) } \nabla \psi dW_{x} +  \hbar k \sqrt{ \langle \sigma_{\textrm{e}} \rangle } \sqrt{2 \gamma E} d\vec{W} \nonumber \\ 
	& + \left( -\tilde g + \frac{\hbar k}{m} \gamma \langle \sigma_{\textrm{e}} \rangle \right)\hat{z} dt,
	\label{eq:Ito_equation_p}
\end{align}

\noindent
where $\Phi = a^{\dagger} \sigma_{\textrm{ge}} + \sigma_{\textrm{eg}} a$, the Wiener increment $dW_{x}^2 = dt$, $d\vec{W}$ is a three-dimensional incremental vector, $E$ denotes the radiation pattern of an atomic dipole ($E_{xx}=2/5, E_{yy} = E_{zz} = 3/10$, and other elements zero), and other terms are

\begin{align}
	\xi(\vec{r}) & = \int^{\infty}_{0} d\tau \left[ \frac{1}{2} \langle \Phi(\tau)\Phi(0) + \Phi(0)\Phi(\tau) \rangle_{\rho_{\textrm{s}}}  -\langle \Phi \rangle^{2}_{\rho_{\textrm{s}}}  \right] \\
	\chi(\vec{r}) &= i\int^{\infty}_{0} d\tau \left[ \tau \langle [\Phi(\tau), \Phi(0)] \rangle_{\rho_{\textrm{s}}} \right],
\end{align}

\noindent
with the steady-state atom-field density matrix $\rho_{\textrm{s}}$.
Outlining the meaning of each term of Eq.~\eqref{eq:Ito_equation_p}, the first one stands for the dipole force by the cavity fields, the cavity-induced friction force is meant by the second one, next term describes the fluctuations of the dipole force, the fourth corresponds to the recoil force caused by the spontaneous emission, and the last term denotes the force induced by gravity and the push beam.
When calculating the trajectory, we numerically integrate the master equation with the Hamiltonian

\begin{align}
	H =  H_{\textrm{int}} + \xi(t)H_{\textrm{ps}},
\end{align}

\noindent
with the step function $\xi(t)=0$ before $t_{\textrm{p}}$ and $\xi(t)=1$ after $t_\textrm{p}$ (push beam turns on at $t_{p}$), and the interaction with the push beam is described by $H_{\textrm{ps}} =  \frac{\Omega_{\textrm{ps}}}{2}(\sigma_{\textrm{ge}} + \textrm{h.~c.})$ ($\Omega_\textrm{ps}$ is the Rabi frequency of the push beam below whose frequency $\omega_{\textrm{ps}} = \omega_{\textrm{p}}$), which provides us with $\rho_\textrm{s}$.
We then calculate $\langle \Phi \rangle$, $\chi(\vec{r})$, and $\xi(\vec{r})$, and $\langle \sigma_{\textrm{e}} \rangle$, determining $d\vec{p}$ via Eq.~\eqref{eq:Ito_equation_p} and the associated atomic position with Eq.~\eqref{eq:Ito_equation_x}.
Complete descriptions are given in Ref.~\cite{Doherty00b}.

\end{document}